\documentclass{ws-ijmpd}
\usepackage[super,compress]{cite}
\usepackage{hyperref}
\hypersetup{colorlinks,urlcolor=black,citecolor=black,linkcolor=black,filecolor=black}

\usepackage{graphicx}
\RequirePackage{subfigure}
\RequirePackage{amsmath}
\begin{document}

\markboth{Y. Huang, D.-J. Liu and X.-Z. Li}{Superradiant instability of $D$-dimensional RN-AdS black hole mirror system}

%
\catchline{}{}{}{}{}
%

\title{Superradiant instability of $D$-dimensional Reissner-Nordstr\"{o}m-anti-de Sitter black hole mirror system}

\author{Yang Huang}

\address{Center for Astrophysics, Shanghai Normal University, 100 Guilin Road, Shanghai 200234, China\\
	saisehuang@163.com}

\author{Dao-Jun Liu}

\address{Center for Astrophysics, Shanghai Normal University, 100 Guilin Road, Shanghai 200234, China\\
	djliu@shnu.edu.cn}

\author{Xin-Zhou Li}
\address{Center for Astrophysics, Shanghai Normal University, 100 Guilin Road, Shanghai 200234, China\\
	kychz@shnu.edu.cn}

\maketitle

\begin{history}
\received{Day Month Year}
\revised{Day Month Year}
\end{history}

\begin{abstract}
	In this paper, a detailed analysis for superradiant stability of the system composed by a $D$-dimensional Reissner-Nordstr\"{o}m-anti-de Sitter (RN-AdS) black hole and a reflecting mirror under charged scalar perturbations are presented in the linear regime. It is found that the stability of the system is heavily affected by the mirror radius as well as the mass of the scalar perturbation, AdS radius and the dimension of space-time. In a higher dimensional space-time, the degree of instability of the superradiant modes will be severely weakened.  Nevertheless,  the degree of instability can be magnified significantly by choosing a suitable value of the mirror radius.  Remarkably, when the mirror radius is smaller than a threshold  value the system becomes stable.  We also find that massive charged scalar fields cannot trigger the instabilities in the background of $D$-dimensional asymptotically flat RN black hole. For a given scalar charge, a small RN-AdS black hole can be superradiantly unstable, while a large one may be always stable under charged scalar field with or without a reflecting mirror.  We also show that these results can be easily expounded and understood with the help of factorized potential analysis.
\end{abstract}


\ccode{PACS numbers:04.70.-s \and 04.50.Gh}

\section{Introduction}
\label{intro}

In classical theory of gravity, black holes are some space-time regions that can not communicate with exterior and no matter and information trapped in the black hole  can  escape to exterior.  Therefore, it is intriguing  that  superradiant scattering, a classical process through which energy can be extracted from a black hole, emerges. For a comprehensive review of superradiance in black hole physics, we refer to  Ref.\refcite{Brito:2015oca} and references therein.

It is well established that bosonic waves impinging on a Kerr black hole can be amplified if the frequency of the wave satisfies the superradiance condition
\begin{equation} 
	\omega<m\Omega_H,
\end{equation} 
where $m$ and $\Omega_H$ are respectively the azimuthal number of the wave mode and the  angular velocity of black hole on the horizon. A similar phenomenon exists in charged Reissner-Nordstr\"{o}m(RN) black hole space-times. In this case, the amplification of an incident charged bosonic wave occurs for the frequency  satisfying
\begin{equation} 
	\omega<q\Phi_H,
\end{equation} 
where $q$ is the charge of the incident bosonic field and $\Phi_H$ is the electric potential at the black hole horizon. More significantly, the superradiant modes may result in the instability of the background if there exists some mechanism via which these modes can be trapped in the vicinity of the black hole \cite{Zeldovich1971}. If a bosonic wave surrounds the black hole with a reflecting mirror, it will bounce back and forth between the mirror and event horizon, amplifying itself each time. Then initially small perturbations could be made to grow without bound until the radiation pressure finally destroys the mirror\cite{PhysRevD.70.044039}. This is the so-called "black hole bomb" ideal proposed by Press and Teukolsky \cite{Press:1972zz}. It is known that the mass of the field itself provides a natural mirror in Kerr space-time \cite{PhysRevD.22.2323,Damour2007}. The instabilities of Kerr black holes under massive bosonic perturbations have been studied extensively \cite{PhysRevD.76.084001,PhysRevD.87.124026,PhysRevD.81.061502,PhysRevD.71.024034,PhysRevD.56.3395,PhysRevLett.84.4537,PhysRevD.70.044018}. 
However,  charged massive bosonic fields cannot result in the instability for RN space-time. Recently, we also show that charged scalar fields can not trigger the instability of black holes in a theory that nonlinear electrodynamics couples to Einstein gravity \cite{Huang:2016}.
In fact, it has been proved by Hod  that four dimensional charged RN black holes are stable to charged massive scalar perturbations in the entire parameter space because the superradiance condition and the bound-state condition which are required in order to trigger the superradiant instability cannot operate simultaneously\cite{Hod2012505,Hod20131489,Hod2015prd}.  Interestingly, it is found that one can still make four dimensional RN black holes unstable to charged scalar fields (massless or massive) by adding a reflecting mirror outside of the black hole \cite{PhysRevD.88.063003,PhysRevD.92.124047}. This frequency-domain conclusion on instability has been confirmed by a time-domain study \cite{PhysRevD.89.063005}. Li \textit{et al.} \cite{0253-6102-63-5-569} further investigate the superradiant instability of the system composed by $D$-dimensional RN black hole, charged massless scalar field, and reflecting mirror outside of the black hole analytically. As a matter of fact, the black-hole-mirror bombs have been studied extensively in the literature \cite{PhysRevD.70.044039,PhysRevD.87.124026,PhysRevD.71.024034,PhysRevD.88.063003,PhysRevD.92.124047,PhysRevD.89.063005,0253-6102-63-5-569,PhysRevD.88.064055,PhysRevD.88.124007,Li:2012nd,Li:2014xxa,Aliev:2014aba,PhysRevD.92.024053,Hod2016177}.

A space-time with a naturally incorporated mirror in it is anti-de Sitter (AdS) space-time, which has attracted a great deal of attention recently due to the so-called AdS/CFT correspondence or gauge/gravity duality \cite{Maldacena:1997re,Gubser:1998,Witten:1998}. In such a duality,  a black hole is dual to a thermal state and a perturbed black hole is dual to a nonthermal boundary gauge theory and the approach to equilibrium in the gravitational side is translated to understanding thermalization in the boundary gauge theory. Especially, some correlation functions and transport coefficients of the dual holographic theory are related the linear response of  black hole in general and the quasinormal frequencies of AdS black holes have a direct interpretation in terms of the dual gauge field theory. Therefore, gaining an insight into the behavior of perturbed black holes in asymptotically AdS space-times is of great relevance for current fundamental and practical researches.

For a black hole in AdS space-time, the black hole bomb effect can be also realized due to the fact that the timelike boundary of the AdS space-time plays the role of a resonant cavity between the black hole and spatial infinity. Therefore, it is not difficult to understand the instability of small Kerr-AdS black holes, as argued in Ref.\refcite{Cardoso:2004hs}\footnote{ The superrandiantly unstable modes of  Kerr-AdS black holes are first explicitly computed in Ref.\refcite{Cardoso:2013pza}.}, although Hawking and Reall have shown that, at least for large Kerr-AdS black holes, this instability is not present \cite{Hawking:1999dp}.  Uchikata and Yoshida \cite{Uchikata:2011zz} explicitly compute the unstable modes of a charged massless scalar field in the background of RN-AdS black holes. Recently, the nonlinear development and the final fate of the  RN-AdS black hole superradiant instability are studied by Bosch, Green and Lehner \cite{Bosch:2016vcp}.   
It was shown that superradiant modes are unstable for Myers-Perry black holes in high-dimensional AdS space-time \cite{Kodama:2009rq}. The superradiant instability for a charged scalar field in a $D$-dimensional small RN-AdS black hole is investigated analytically and numerically  in frequency domain by Wang and Herdeiro \cite{PhysRevD.89.084062}. Superradiance and instability of small rotating charged AdS black holes in all dimensions are studied by Aliev \cite{Aliev:2016}.   Recently, the time evolution of scalar field perturbations in $D$-dimensional RN-AdS black holes is also investigated \cite{Li:2016kws}.\footnote{In addition to scalar perturbations,  the stability against gravitational perturbations, i.e. the stability of the background itself  has been  shown for RN-AdS black holes \cite{Konoplya:2008rq}. Ref. \refcite{Konoplya:2011qq} is a useful review which also discuss superradiant instabilities.}

The main purpose of the present paper is to investigate the superradiant instability of the system composed by a $D$-dimensional charged RN black hole in AdS space-time and a reflecting mirror outside of the black hole. 
The paper is organized as follows: In Sec. \ref{sec:background}, we introduce the background space-time and the charged scalar perturbation equation with mirror-like boundary condition which is numerically explored in Sec. \ref{sec:results}. After an explanation by using factorized potential analysis in Sec. \ref{sec:discussion}, we draw some conclusions in the last section. Throughout the paper, we use {natural units} in which $G=c=\hbar=1$.

\section{Background and perturbation equations}\label{sec:background}
We shall consider a charged, massive scalar field $\Psi$ propagating in the background of a $D$-dimensional RN-AdS black hole. In Boyer-Lindquist-type coordinates, the space-time outside the black hole is described by the following line element
\begin{equation}
	ds^2=-f(r)dt^2+\frac{dr^2}{f(r)}+r^2d\Omega^2_{D-2},
\end{equation}
where $d\Omega^2_{D-2}$ denotes the line element of the $(D-2)$-dimensional unit sphere. The metric function is given by
\begin{equation}\label{func: f(r)}
	f(r)=1-\frac{2M}{r^{D-3}}+\frac{Q^2}{r^{2(D-3)}}+\lambda r^2.
\end{equation}
Here, the parameter $\lambda$ is related to the cosmological constant in the following way
\begin{equation}
	\lambda=-\frac{2\Lambda}{(D-2)(D-1)}.
\end{equation}
Note that $\lambda$ is positive for AdS space-time, therefore it will be convenient to define $L^2=1/\lambda$ where $L$ is called AdS radius. The quantities $M$ and $Q$ in Eq.(\ref{func: f(r)}) are respectively given by the mass $\tilde{M}$ and the charge $\tilde{Q}$ of the black hole via \cite{Kodama01012004}
\begin{equation}
	M=\frac{8\pi\tilde{M}}{(D-2)V_{D-2}},\quad Q^2=\frac{8\pi\tilde{Q}^2}{(D-2)(D-3)},
\end{equation}
where the volume of the $(D-2)$-sphere is $V_{D-2}=\frac{2\pi^{\frac{D-1}{2}}}{\Gamma(\frac{D-1}{2})}$. It is noted that, just as RN black hole in four dimensional space-time,  the higher-dimensional RN-AdS black hole usually has two horizons:  the event horizon at $r=r_+$, which we consider it as the edge of the black hole and the Cauchy horizon at $r=r_-$ inside the black hole. For convenience, we measure all quantities in terms of the event horizon $r_+$ and set $r_+=1$ throughout the paper. According to the definition of the horizon, we have
\begin{equation}
	f(r_-)=f(r_+=1)=0,
\end{equation}
then it can be obtained that
\begin{equation}\label{black hole mass and charge}
	\begin{aligned}
		&M=\frac{-(\lambda+1)r_-^2+\lambda r_-^{2 D-2}+r_-^{2D-4}}{2\left(r_-^{D-1}-r_-^2\right)},\\
		&Q^2=\frac{-(\lambda+1)r_-^D +\lambda r_-^{2D-1}+r_-^{2D-3}}{r_-^{D}-r_-^{3}}.
	\end{aligned}
\end{equation}

For non-extremal black holes, we have $r_-<r_+$, correspondingly, $Q<Q_c$, where the critical charge $Q_c$ corresponds to  the maximal charge which,  in our units,  is determined by 
\begin{equation}\label{extrem charge}
	Q^2_c=1+\frac{D-1}{D-3}\lambda, 
\end{equation}
for given values of $\lambda$ and $D$.
On the other hand, for extremal black holes, $r_-=r_+=1$ and $Q=Q_c$. Due to the spherical symmetry of the background space-time, we can set the angular components of the electromagnetic potential of the black hole  to zero and select a gauge to let the radial component vanish, then the electromagnetic potential is 
\begin{equation}\label{Pot:the electromagnetic potential}
	A_{\mu}dx^{\mu}=-\Phi(r)dt=-\sqrt{\frac{D-2}{2(D-3)}}\frac{Q}{r^{D-3}}dt,
\end{equation}
where we have taken a vanishing electromagnetic potential at spatial infinity.

Neglecting the back-reaction of the scalar field on the metric and on the electromagnetic field, we can describe the dynamics of the charged scalar field by the following Klein-Gordon equation
\begin{equation}\label{Eq: Klein Gordon equation}
	(\nabla_{\mu}-iqA_{\mu})(\nabla^{\mu}-iqA^{\mu})\Psi=\mu^2\Psi,
\end{equation}
where  $\nabla_{\mu}$ is the covariant derivatives in RN-AdS space-time and the vector potential $A_{\mu}$ is determined by Eq.(\ref{Pot:the electromagnetic potential}). Here, $q$ and $\mu$ are the charge and mass of the field, respectively.    It is convenient to decompose the scalar field in the form
\begin{equation}\label{decomposition ansatz} \Psi(t,r,\theta_i,\phi)=r^{-(D-2)/2}\psi_{\ell}(r)Y_{\ell,D-2}(\theta_i,\phi)e^{-i\omega t} \end{equation}
where $Y_{\ell,D-2}(\theta_i,\phi)$ denotes the generalized spherical harmonics on the $(D-2)$-sphere. Here $\ell$ is spherical harmonic index and the azimuthal harmonic index is omitted.  Substituting Eq.(\ref{decomposition ansatz}) into the Klein-Gordon equation (\ref{Eq: Klein Gordon equation}), we obtain  the  equation for radial function $\psi_{\ell}(r)$
\begin{equation}\label{Eq: decomposed equation}
	f^2(r)\frac{d^2\psi_{\ell}}{dr^2}+f(r)f'(r)\frac{d\psi_{\ell}}{dr}+(\omega-q\Phi(r))^2\psi_{\ell}-V(r)\psi_{\ell}=0,
\end{equation}
where $\Phi(r)$ is given in Eq.(\ref{Pot:the electromagnetic potential}) and the potential function 
\begin{eqnarray}\label{V}
	V(r)&&=\frac{(D-2)(D-4)f^2(r)}{4r^2}\nonumber\\
	&&+f(r)\left[\frac{(D-2)f'(r)}{2r}+\frac{\ell(\ell+D-3)}{r^2}+\mu^2\right].
\end{eqnarray}
Since Eq.(\ref{Eq: decomposed equation}) is unchanged under $\omega\rightarrow-\omega$ and $q\Phi(r)\rightarrow-q\Phi(r)$,  we consider only the case for $\omega>0$ and $q\Phi(r)>0$ without loss of generality.
Eq.(\ref{Eq: decomposed equation}) can be further rewritten in  the form
\begin{equation}\label{eq:wave13}
	\left[\frac{d^2}{dr_*^2}+\left(\omega-q\Phi(r)\right)^2-V(r)\right]\psi_{\ell}=0,
\end{equation}
Here, we have introduced the tortoise coordinate $r_*$, which is defined by $r_*=\int\frac{dr}{f(r)}$.

The characteristic equation (\ref{Eq: decomposed equation}) for the radial function $\psi_{\ell}(r)$ should be supplemented by appropriate boundary conditions.  Near the event horizon $r_+$, an ingoing wavelike condition is imposed as usual
\begin{equation}\label{cd: Ingoing wavelike condition at horizon-1}
	\psi_{\ell}(r)\sim e^{-i(\omega-\omega_c)r_*}\;\;\mathrm{as}\;\; r\rightarrow r_+ \ (r_\ast\rightarrow -\infty),
\end{equation}
where the critical frequency $\omega_c=q\Phi(r_+)\equiv q\Phi_H$. 
One can find that if $\omega <\omega_c$, the mode appears to be outgoing for an inertial observer at spatial infinity. This is just the superradiant condition of scalar field in $D$-dimensional RN black hole. 

In addition, we impose a  perfect reflecting "mirror" boundary condition, that is,  the scalar field vanishes at the mirror's location $r_m$,  
\begin{equation}\label{eq:mirrorCondition}
	\psi_{\ell}(r=r_m)=0.
\end{equation}
The characteristic radial equation (\ref{Eq: decomposed equation}) together with above boundary conditions forms a two-point boundary value problem, which  will lead to a discrete spectrum of modes with complex frequencies 
\begin{equation}
	\omega=\omega_R+i\ \omega_I, 
\end{equation}
where $\omega_R$ and $\omega_I$ are two real quantities denoting the real part and imaginary part of frequency, respectively.
Since the scalar field has the time dependence $\Psi\sim e^{-i\omega t}$, $\omega_I>0$ or  $\omega_I<0$ implies the amplitude of the mode exponentially grows or decays, respectively. 

\section{Numerical Analysis}\label{sec:results}
\begin{figure*}    
	\subfigure { \label{fig:7a}     
		\includegraphics[width=0.45\textwidth]{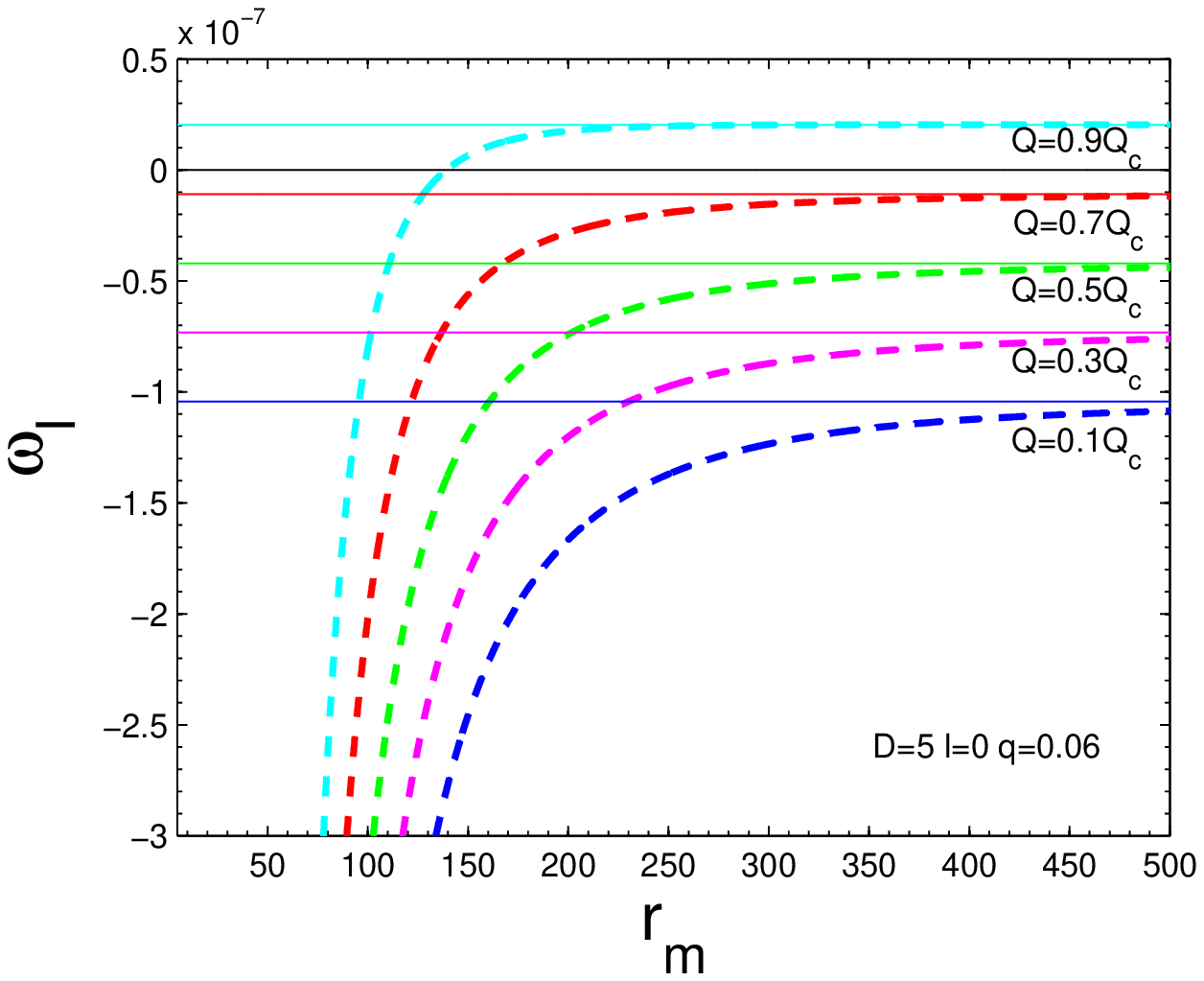}  }
	\subfigure { \label{fig:7b}     
		\includegraphics[width=0.45\textwidth]{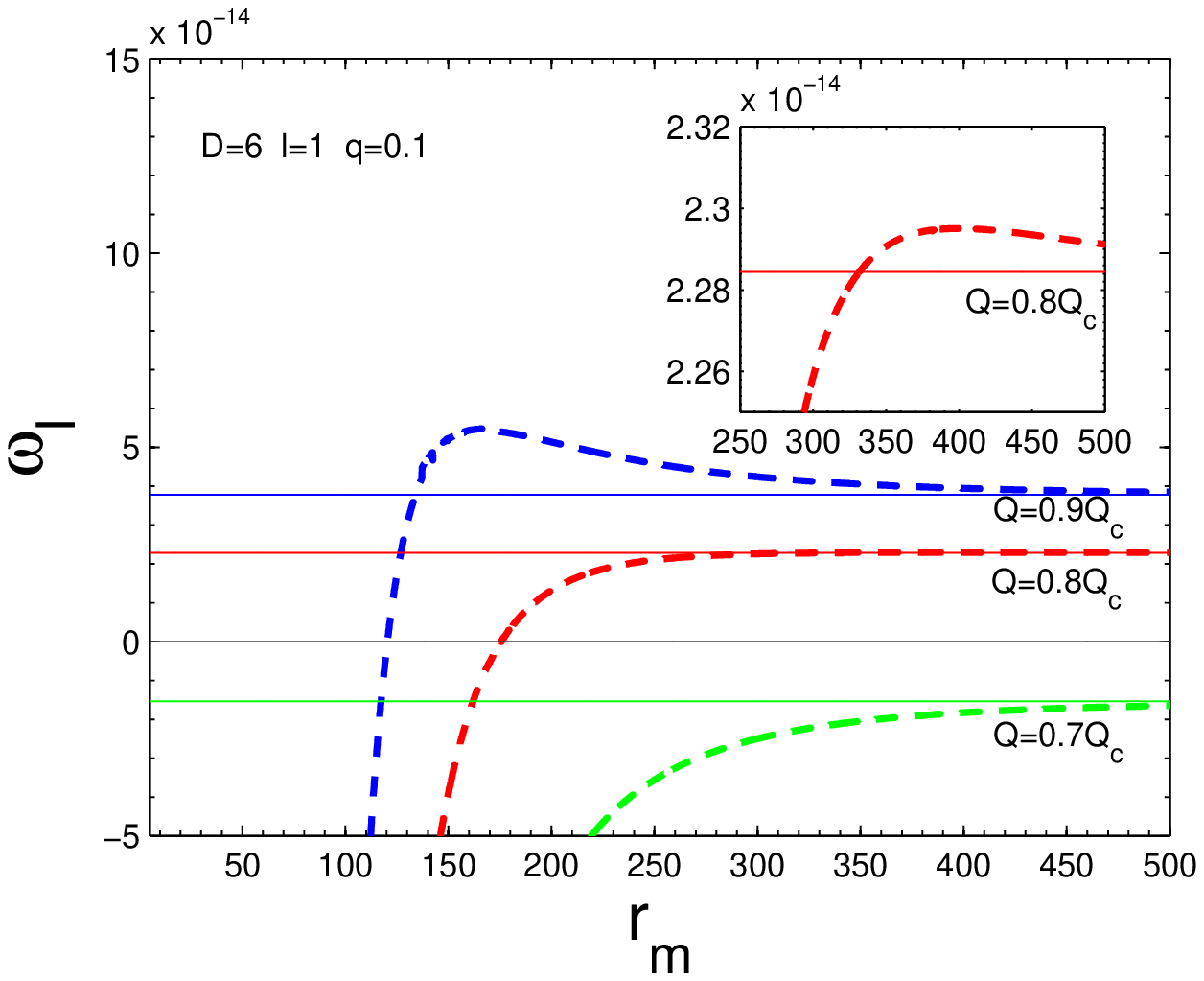}  }
	\caption{The dashed lines are the numerical results of the imaginary part of frequency of the fundamental modes as a function of the mirror radius for various values of the charge $Q$ of a higher dimensional RN-AdS black hole and  the solid horizontal lines are from the results reported in Ref.\cite{PhysRevD.89.084062}. In the left panel, we take the parameters $D=5,\ \ell=0,\ q=0.06,\ \mu=0.0,$ and $ \lambda=0.0001$, which are corresponding to those in table II of Ref.\cite{PhysRevD.89.084062}, while in the right panel, the parameters are chosen to be $D=6,\ \ell=1,\  q=0.1,\ \mu=0.0,$ and $\lambda=0.0001$ corresponding to those in table III of Ref.\cite{PhysRevD.89.084062}. }
	\label{fig:7}
\end{figure*}

To compute the spectrum of bound-state modes numerically, we use an algorithm similar to the one adopted in Ref. \refcite{PhysRevD.88.063003}. We start integrating the radial equation with the behavior (\ref{cd: Ingoing wavelike condition at horizon-1}) from $r=r_+(1+\varepsilon)$ ($\varepsilon=10^{-5}$) to $r_m$ with an initial value of complex frequency $\omega$. This will give us a value for the wave function at $r_m$. Obviously, it is dependent on the value of the complex frequency $\omega$. Changing the value of the frequency and repeating the integration procedure until the boundary condition (\ref{eq:mirrorCondition}) is satisfied with the desired precision, we obtain the frequency of the bound state. 

What follows is a summarization of  our numerical results. 
We show in Fig.\ref{fig:7} the imaginary part of frequency $\omega_I$ of the fundamental mode of a typical charged scalar perturbation  as a function of the mirror radius for various values of a higher-dimensional RN-AdS  black hole's charge and compare them with the results reported in Ref.\cite{PhysRevD.89.084062}. 
It is clear that, if a $D$-dimensional RN-AdS black hole is  unstable under a given charged perturbation\footnote{This can be seen from the value of $\omega_I$ when $r_m$ goes to infinity.}, surrounding it with a reflecting mirror may change its stability, which is dependent of the mirror's radius. In other words, to keep a positive value of $\omega_I$,  the mirror's radius should be greater than a threshold value which is also dependent of the parameters of the black hole. On the contrary, if such a RN-AdS black hole is stable against a specific charged perturbation, its stability will not change qualitatively in the presence of a mirror. In this situation, $\omega_I$ is alway smaller than $0$, although it grows monotonically as $r_m$ increases.
On the other hand, the potential function (\ref{V}) is divergent at spatial infinity in AdS space-time, which implies $\psi_{\ell}(r=\infty)=0$, therefore, as the mirror is moved away from the black hole, our numerical results should reduce to the analytic results reported in Ref.\refcite{PhysRevD.89.084062} in some degenerate cases. Indeed, from Fig.\ref{fig:7}, it is obvious that when the mirror radius is large enough ($r_m \simeq 500r_+$ in the figure), they are all in good agreements\footnote{Note that, in Ref. \refcite{PhysRevD.89.084062} all physical quantities are normalized by the AdS radius $L=1/\sqrt{\lambda}$ and event horizon $r_+$ is set to be $0.01$. However, we measure all the quantities in terms of the event horizon and set $r_+=1$ here. Thus, there is a difference of $100$ times between the numerical value of our results and those in Ref. \refcite{PhysRevD.89.084062}.}. This may be considered as a confirmation for the validity of our codes. 

\begin{figure*}    
	\subfigure { \label{fig:2a}     
		\includegraphics[width=0.45\textwidth]{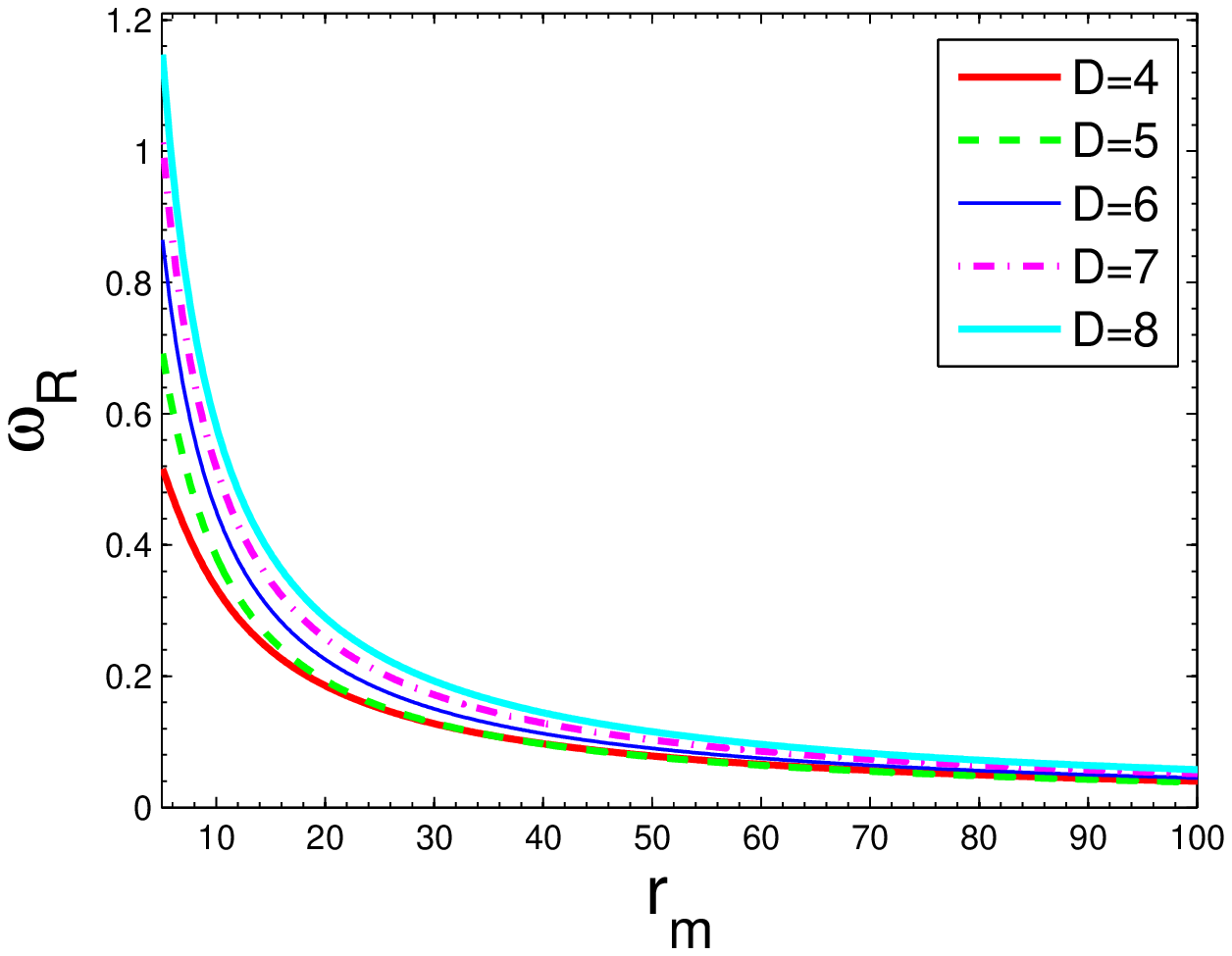} 
	}
	\subfigure { \label{fig:2b}     
		\includegraphics[width=0.45\textwidth]{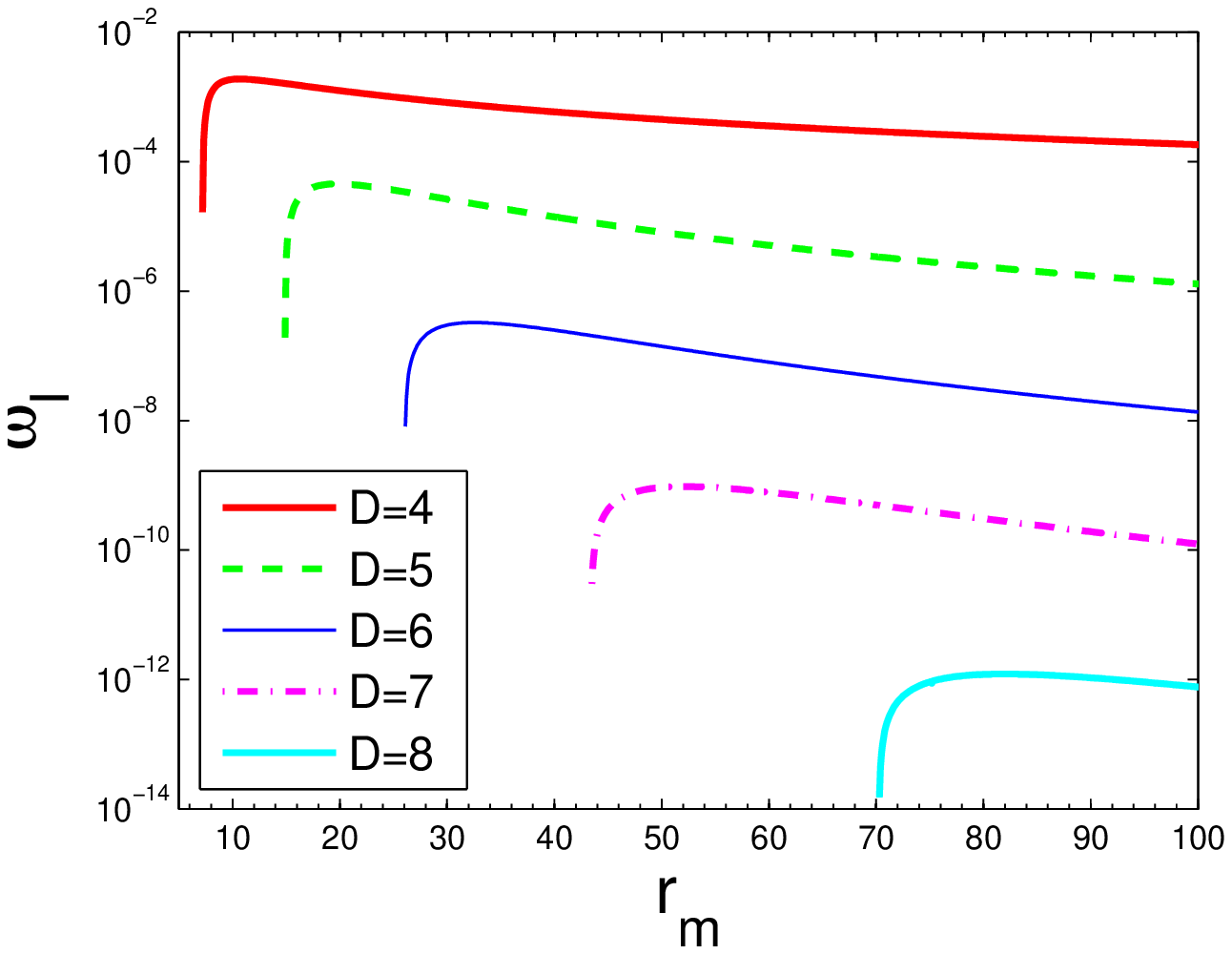} 
	}
	\caption{The real (left) and imaginary (right) part of the fundamental frequencies $\omega$ of charged massless scalar perturbations, with $q=0.6$ and $\ell=0$, as a function of the mirror radius $r_m$ in different dimensional RN ($\lambda=0.0$) space-times, where we have set $r_-=0.5$.}
	\label{fig:2}     
\end{figure*}

What we are interested in is the influence of physical parameters (such as Cauchy horizon of black hole $r_-$, the dimension of space-time $D$, the mass of perturbation $\mu$ and AdS radius $L$) on the stability of the black hole-mirror system.

To this end, we first plot the fundamental frequency $\omega$ of s-wave ($\ell=0$) of a charged massless scalar perturbation, as a function of the mirror radius $r_m$ in some different dimensional RN ($\lambda=0.0$) space-times in Fig.\ref{fig:2}. Clearly, the real part does not shift significantly with the change of the space-time's dimension while the imaginary part changes enormously. Moreover, the higher the dimension of the space-time, the smaller the value of the imaginary part of the frequency. It is noteworthy that by solving the system (\ref{Eq: decomposed equation}) subjected to boundary conditions (\ref{cd: Ingoing wavelike condition at horizon-1}) and (\ref{eq:mirrorCondition}), we can in principle compute the spectrum of bound states in arbitrary dimensional space-times, but when the dimension of space-time is so high that one would obtain a very small imaginary part of the frequency (e.g., the case of $D=8$ in Fig.\ref{fig:2}, $\omega_I\sim10^{-16}$ for some value of mirror radius). It becomes difficult to numerically compute such a small quantity due to the limited numerical precision. Therefore, without loss of generality,  our focus is mainly on the case of five or six dimensional space-times. 

\begin{figure*}
	\subfigure { \label{fig:1b}     
		\includegraphics[width=0.45\textwidth]{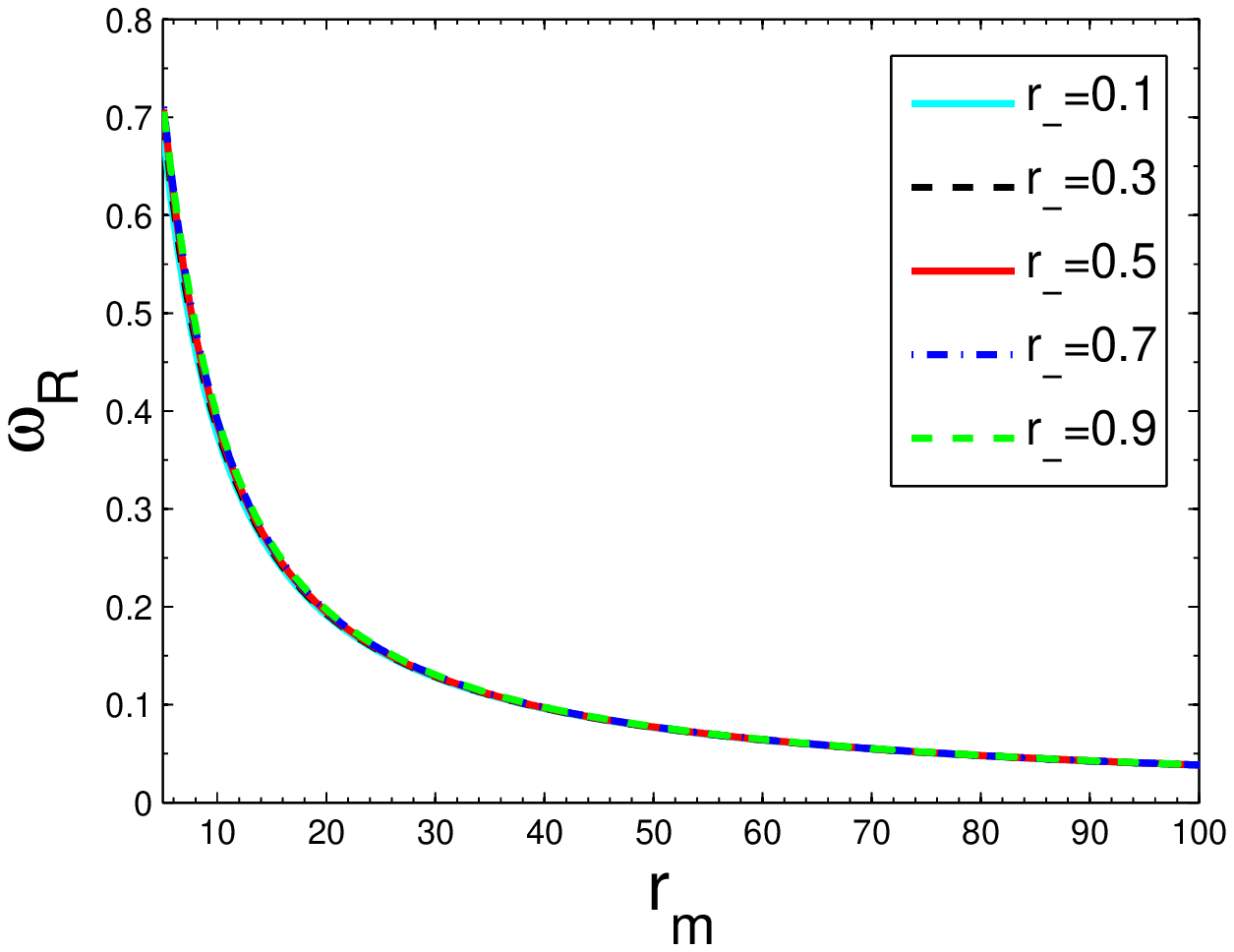} 
	}
	\subfigure { \label{fig:1a}     
		\includegraphics[width=0.45\textwidth]{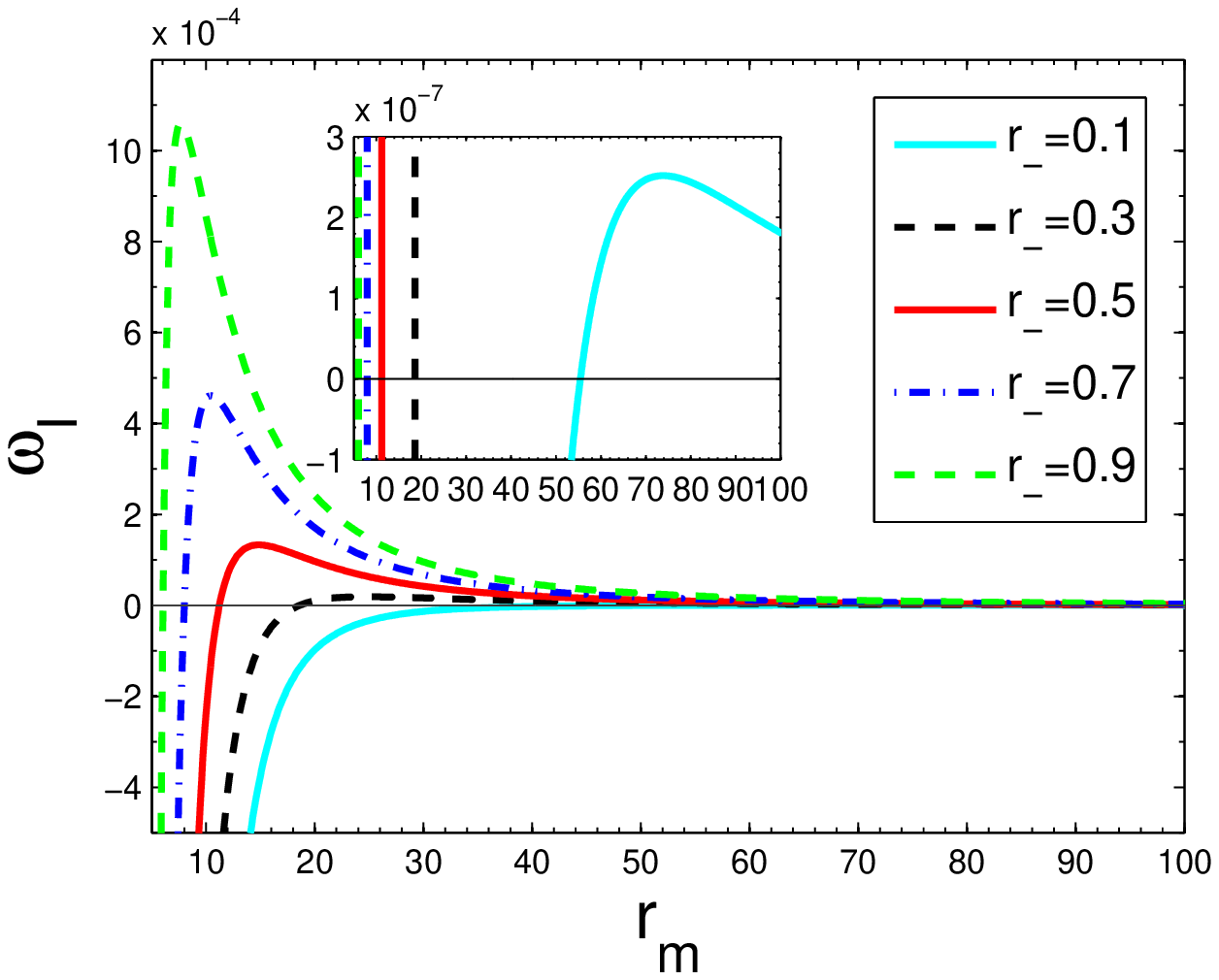} 
	}
	\caption{The real and imaginary part of the fundamental frequency $\omega$ of the mirrored quasibound states, with $\ell=0,\ q=0.8$ and $\mu=0.0$, as a function of the mirror radius $r_m$ in a five dimensional RN space-time with different values of the black hole Cauchy horizon $r_-$.}
	\label{fig:1}
\end{figure*}

\begin{figure*}    
	\subfigure { \label{fig:3a}     
		\includegraphics[width=0.45\textwidth]{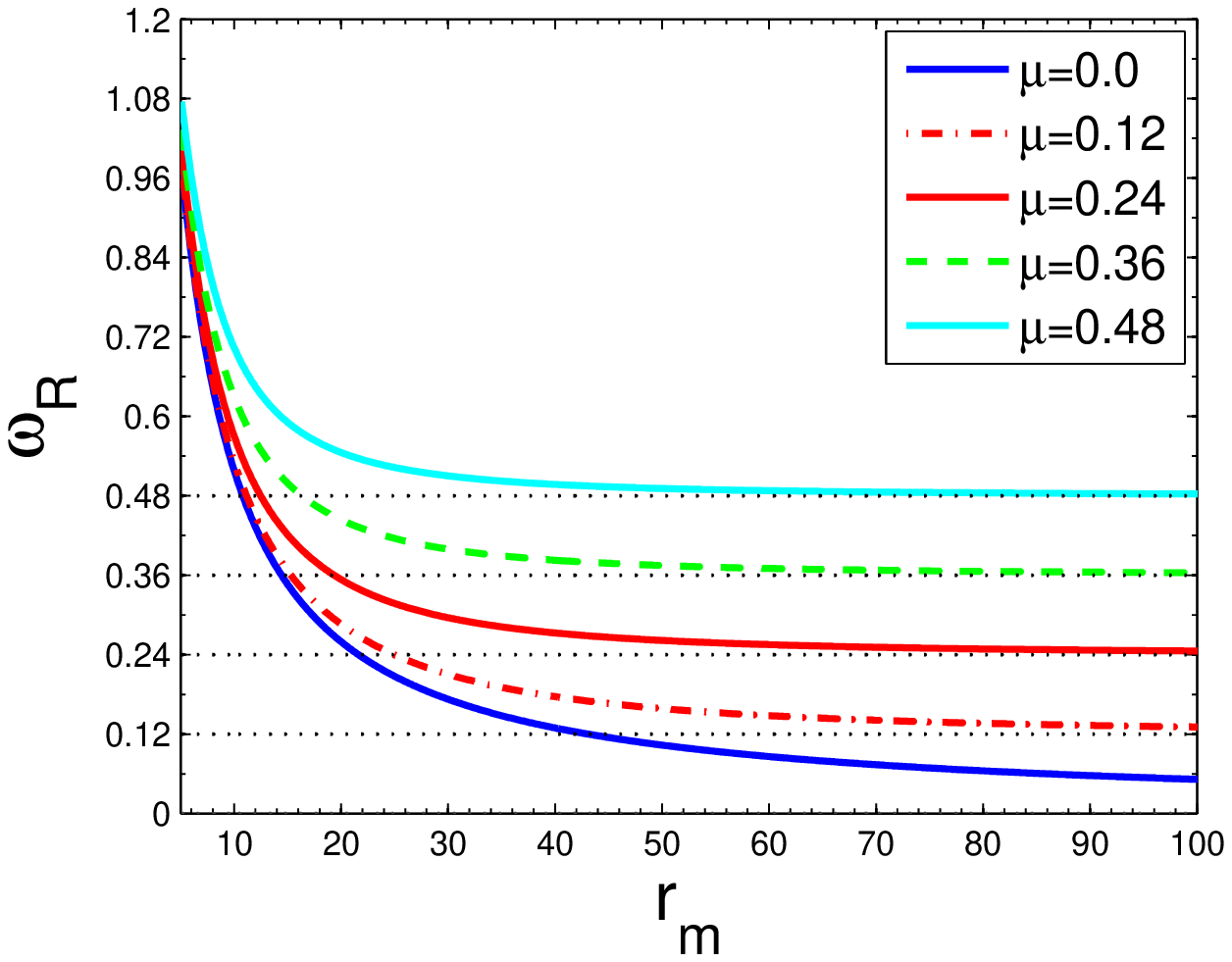}	}
	\subfigure { \label{fig:3b}     
		\includegraphics[width=0.45\textwidth]{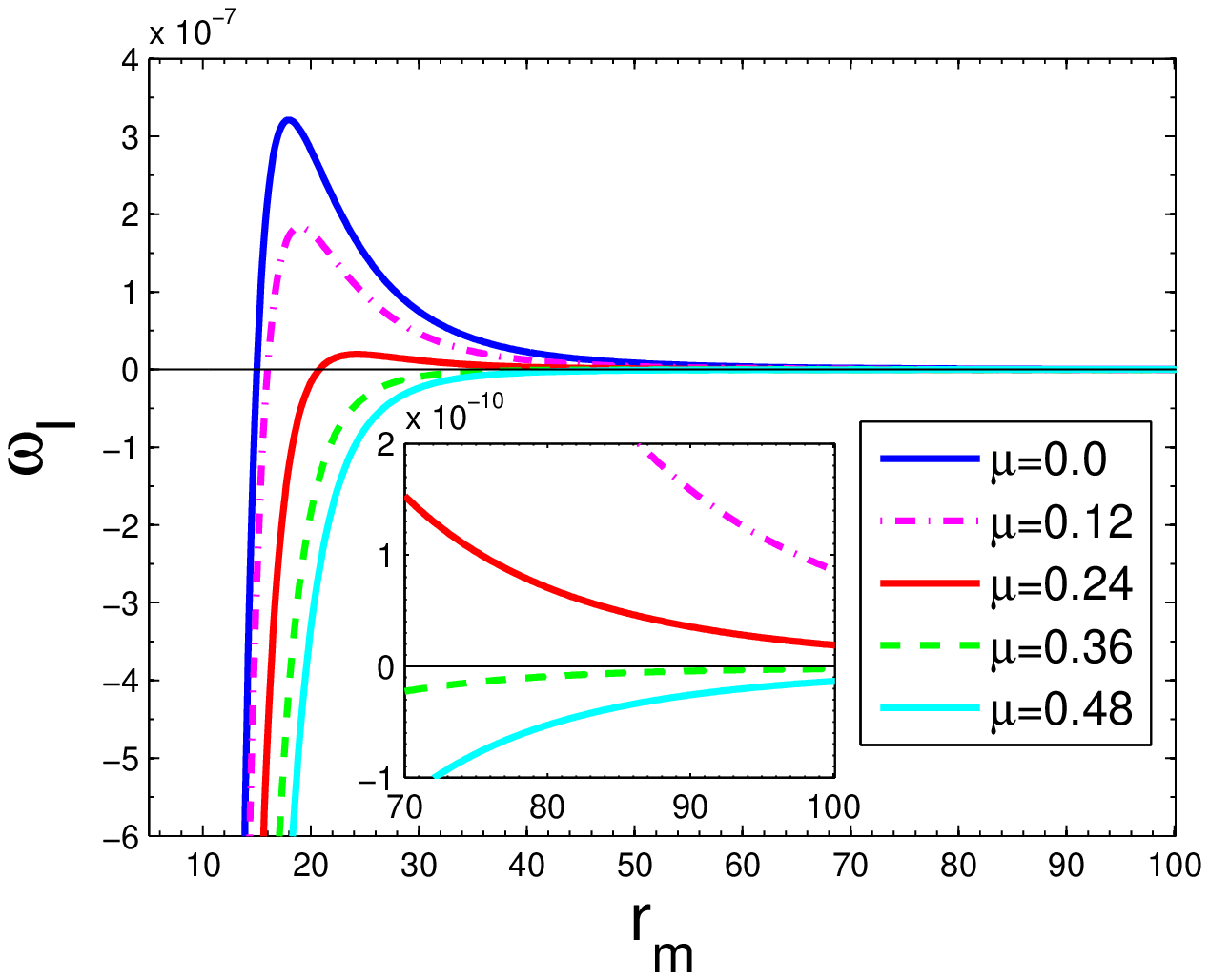} }
	
	\caption{ The frequency of mirrored quasi-bound modes of charged scalar perturbations with different values of mass $\mu$  as a function of the mirror radius $r_m$ in a five dimensional RN space-time. The dotted horizontal straight lines label the values of mass. Here we set $\ell=1$, $q=4.0$ and $r_-=0.1$.
	}
	\label{fig:3}     
\end{figure*}

Next, in Fig.\ref{fig:1}, we  show the real and imaginary part of the fundamental frequency $\omega$ of the mirrored quasi-bound modes, with $\ell=0,\ q=0.8$ and $\mu=0.0$, as a function of the mirror radius $r_m$ in a five dimensional RN space-time with different values of the black hole Cauchy horizon $r_-$. Obviously, the real part of the frequency $\omega_R$ decreases monotonically with $r_m$ and  $r_-$ has a negligible effect on $\omega_R$. However, $r_-$ can affect drastically the imaginary part of frequency $\omega_I$. 
It is shown  that the mirror radius need to be greater than a threshold  value to have a positive imaginary part of the frequency and the greater $r_{-}$ becomes, the smaller the threshold  value of $r_m$ takes. Furthermore,  $\omega_I$ has a positive maximum  value for some value of $r_m$ and reduces to zero at infinity.  For different values of $r_{-}$, the maximum value of $\omega_I$ can differ by orders of magnitude. 
It should be pointed out that although the results presented here are obtained in the five dimensional space-time, similar results should be also obtained in higher dimensional space-times as well as in four dimensional space-time \cite{PhysRevD.88.063003,PhysRevD.92.124047}.

Now, we would like to investigate the influence of the perturbation's mass on the stability. In Fig.\ref{fig:3}, we plot the real and imaginary part of the frequency of  quasi-bound modes of charged scalar perturbations with different values of mass $\mu$ in a five dimensional RN space-time as a function of mirror radius $r_m$. 
Clearly, the real part of the frequency $\omega_R$ decreases monotonically with $r_m$ and reduces to the value of $\mu$ as the mirror is moved far away from the black hole. From the right panel, we can find that in the RN black hole mirror system, the greater the value of  mass, the smaller the value of $\omega_I$. When the mass is greater than some value, the unstable modes would disappear. These results are consistent with those of Ref.\refcite{PhysRevD.88.063003}.

\begin{figure*}    
	\subfigure { \label{fig:3c}     
		\includegraphics[width=0.45\textwidth]{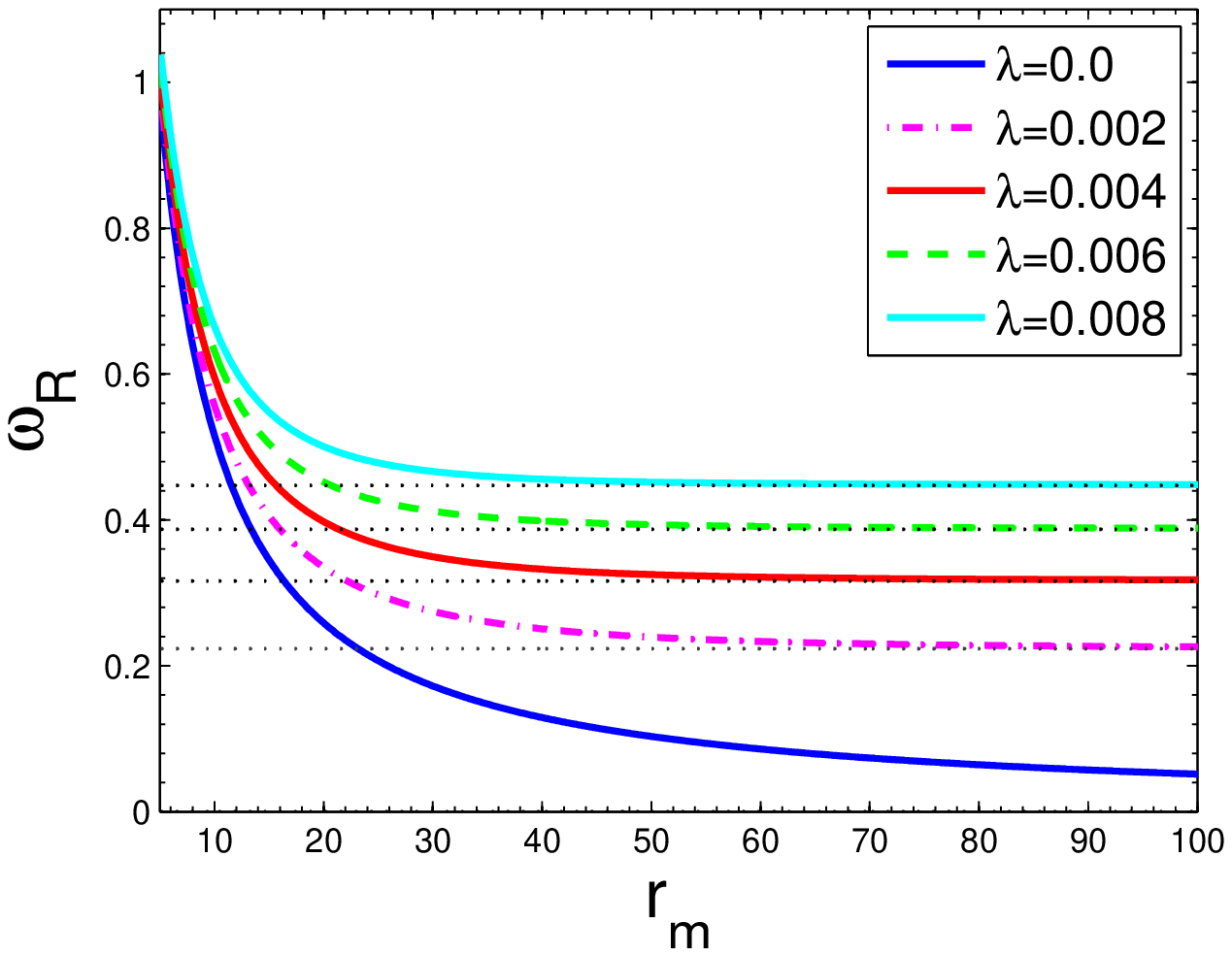} 
	}
	\subfigure { \label{fig:3d}     
		\includegraphics[width=0.45\textwidth]{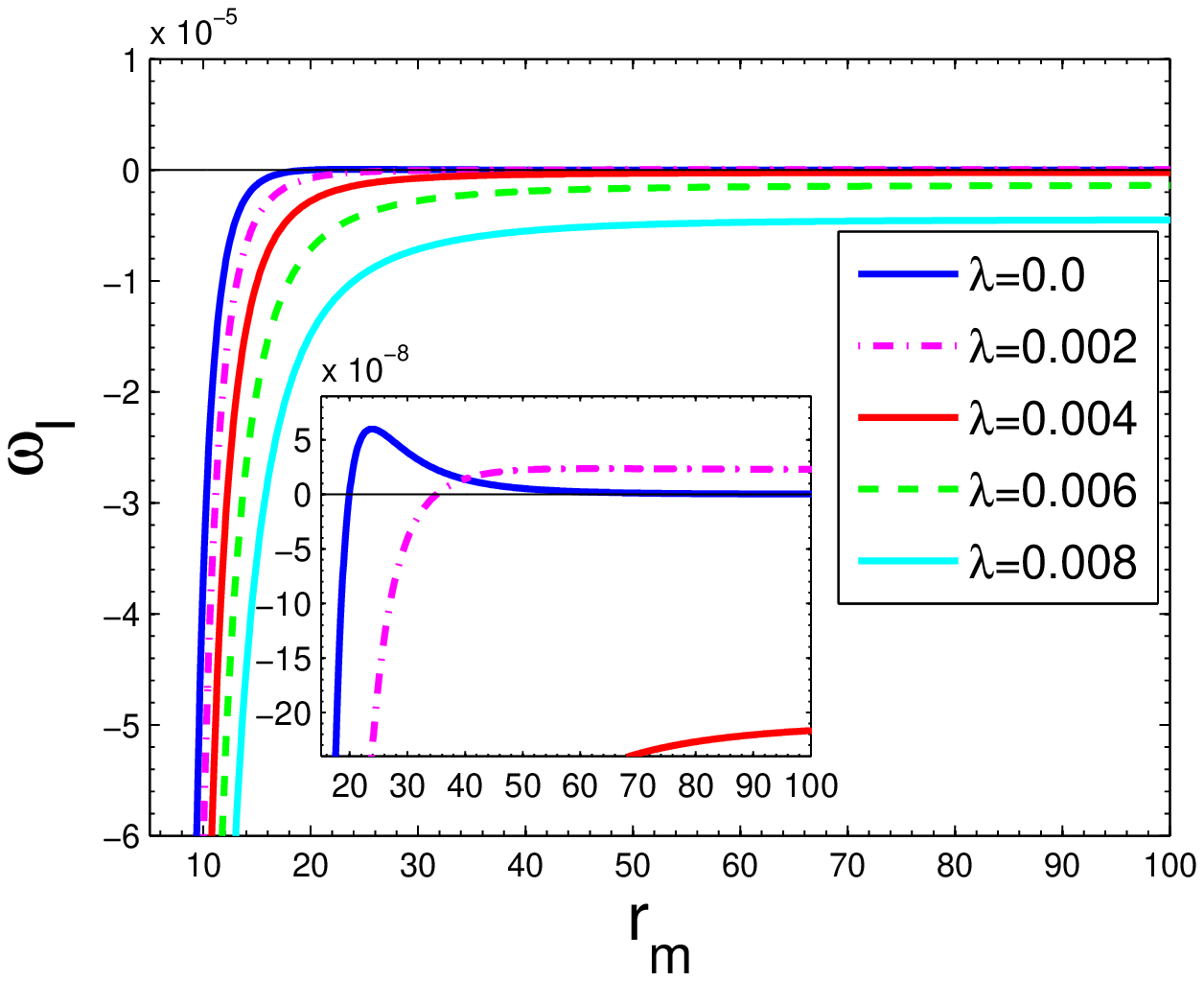}	
	}
	\caption{ The frequency of mirrored quasi-bound modes of a charged massless scalar perturbation as a function of the mirror radius $r_m$ in five dimensional RN-AdS space-times with different values of $\lambda$.  Here, we set $\ell=1$, $q=3.0$ and $r_-=0.1$.
	}
	\label{fig:5-1}     
\end{figure*}

Finally, we study the effect of parameter $\lambda$ (equivalently, cosmological constant or AdS radius) on the stability of the black hole mirror system. For this, we plot the frequency of mirrored quasi-bound modes of a charged massless scalar perturbation as a function of the mirror radius $r_m$ in five dimensional RN-AdS space-times with different values of $\lambda$ in Fig.\ref{fig:5-1}. Remarkably, the effect of $\lambda$ on the real part of the frequency $\omega_R$ is very similar to that of perturbation's mass. The nonzero cosmological constant or $\lambda$ will increase the value of $\omega_R$, although this effect is not noticeable if the mirror radius is small. As the mirror goes to infinity, the asymptotic value of $\omega_R$ is given by the AdS spectrum \cite{PhysRevD.89.084062}
\begin{equation}\label{AdS spectrum} 
	\omega_{\mathrm{AdS}}=(2N+\ell+D-1)\sqrt{\lambda}, 
\end{equation}
where $N$ is the overtone number of the mode. Note that in Fig.\ref{fig:5-1}, only fundamental modes (i.e., $N=0$) are plotted.

However, the effect of $\lambda$ on the imaginary part of the frequency $\omega_I$ is different from that of perturbation's mass. From the right panel, we realize that for the case $\lambda\neq 0$, $\omega_I$ does not vanishes when the reflecting mirror is moved far away from the black hole. Besides, for a given value of charge of perturbation, there exists a threshold  value for $r_m$ when the value of $\lambda$ is relatively small. When $r_m$ is greater than this threshold  value, $\omega_I$ becomes positive, which means the corresponding modes become unstable. 
The larger $\lambda$ is, the greater the threshold  value of $r_m$ takes. Furthermore, the imaginary part $\omega_I$ has a maximum and it decreases to a non-vanishing constant  as the mirror moves  to spatial infinity. However, when the value of $\lambda$ is relatively large ($\lambda=0.008$ for example), $\omega_I$ is always negative, no matter how large the value of mirror radius is. In this circumstance, the mirror will make the system more stable.
It is worthwhile to note that, a greater value of $\lambda$ corresponds to a smaller AdS radius $L$, and since we measure the quantities in terms of event horizon $r_+$, a greater value of $\lambda$ actually means a relatively larger black hole measured by AdS radius.  Therefore, the above result is equivalent to tell us that, for a given charged scalar perturbation,  small RN-AdS black hole surrounded by a mirror can present superradiant instability, but a larger one may be always stable. Of course,  how large a black hole can be said a large black hole in this context depends on the perturbation's charge $q$. Since there is no bound on $q$ at least on the classical level, a large RN-AdS black hole may be also unstable if the charge is large enough. We will see this more clearly in the next section.

\section{Factorized potential and its application to the stability analysis}\label{sec:discussion}
To understand our numerical results presented above and make our statements more concrete, it is useful to perform a factorized potential analysis \cite{1126-6708-2005-07-009}.

\subsection{Factorized potential}
The factorized potential analysis is effective in the qualitative investigation of the Schr\"{o}dinger-like equation.
For this purpose, we rewrite radial equation (\ref{eq:wave13}) as
\begin{equation}
	\frac{d^2\psi}{dr_*^2}+(\omega-V_+)(\omega-V_-)\psi=0,
\end{equation}
where the factorized potentials
\begin{equation}\label{Pot: factorized potential}
	V_{\pm}=q\Phi(r)\pm\sqrt{V(r)}.
\end{equation}
Here $\Phi(r)$ and $V(r)$ are defined in Eq.(\ref{Pot:the electromagnetic potential}) and Eq.(\ref{V}), respectively. Since $V(r)$ is positive definite outside the event horizon, we have $\quad V_+\geq V_-$. Clearly, the field $\psi$ has propagative character when $\omega>V_{+}｝$ or $\omega<V_{-}$ (classically allowed) and evanescent character when $V_-<\omega<V_+$ (classically forbidden). For a given mode, to form a bound state, its propagative zone needs to be   sandwiched between two evanescent zones. On the other hand, from Eq.(\ref{Pot: factorized potential}) we have $V_+(r_+)=q\Phi_H=\omega_c$, hence, a superradiant scattering occurs only when there are some regions  satisfying the condition $V_+(r)<V_+(r_+)=\omega_c$. We call these regions "superradiance zones".

\begin{figure*}    
	\subfigure { \label{fig:4a}     
		\includegraphics[width=0.45\textwidth]{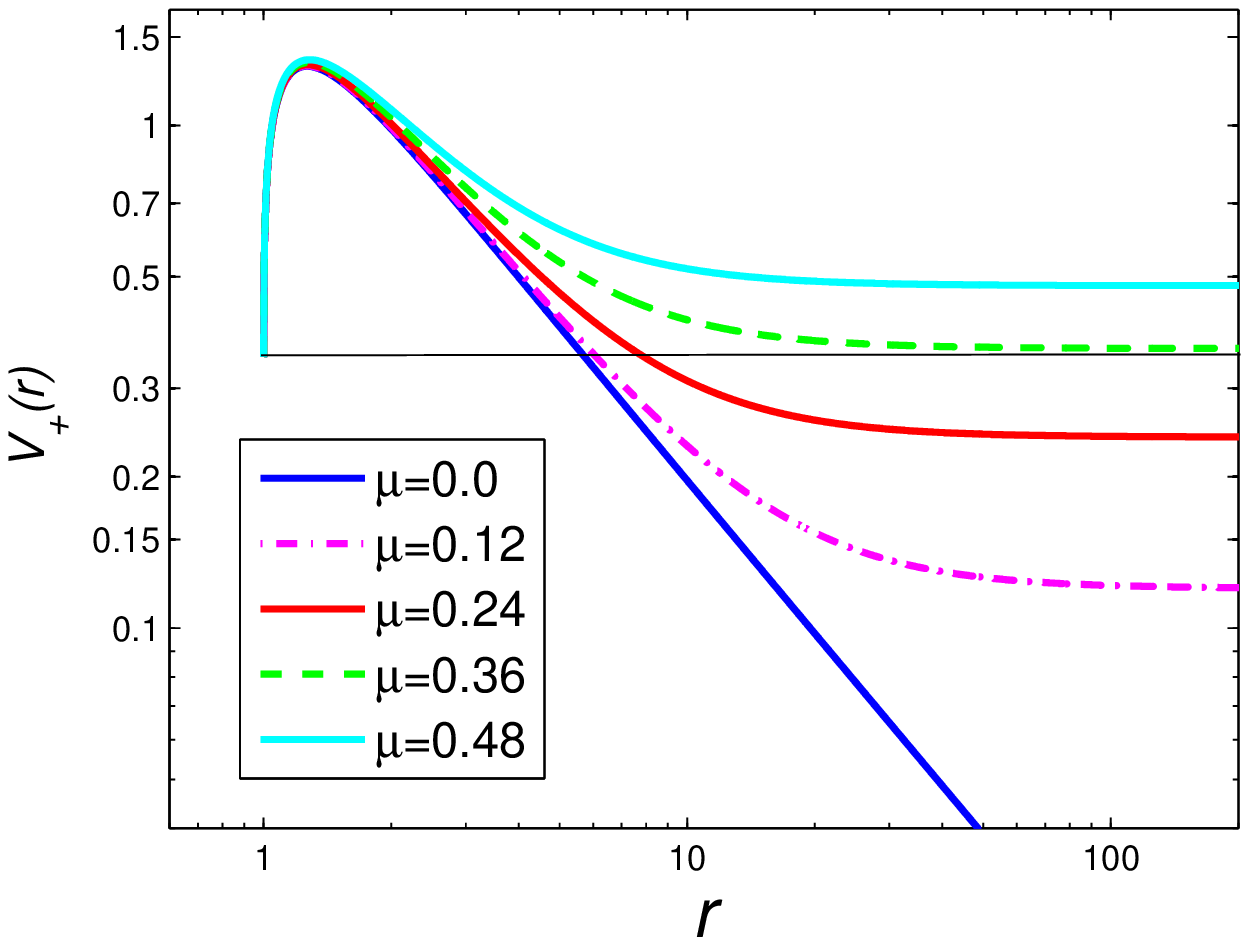} 
	}
	\subfigure { \label{fig:4b}     
		\includegraphics[width=0.45\textwidth]{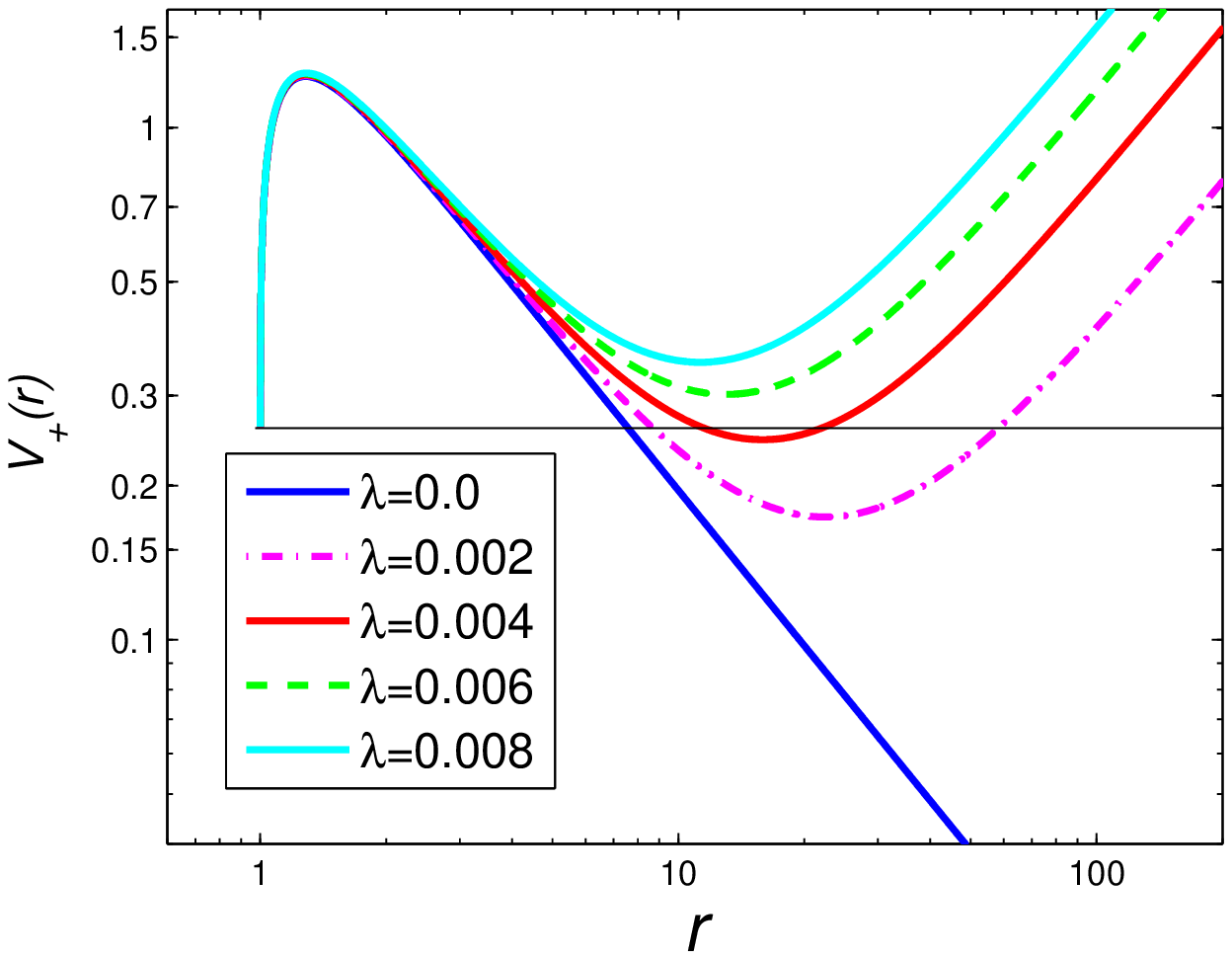}
	}
	\caption{The factorized potentials $V_+(r)$ in five dimensional space-time are plotted by using double logarithmic coordinates. Left panel: $q=4.0$,  $\lambda=0.0$ and the mass of the scalar perturbation $\mu=0.0, 0.12,0.24,0.36,0.48$; Right panel: $q=3.0$, $\mu=0.0$ and  $\lambda=0.0, 0.002,0.004,0.006, 0.008$. In both panels, we set $r_-=0.1$ and the solid horizontal line denotes the critical value $\omega_c$.}
	\label{fig:4}     
\end{figure*}

As is mentioned in Sec.\ref{sec:background},  we restrict our discussions for the case of $\omega>0$. Since the possible bound states for $\omega>0$ are determined by $V_{+}$, we focus on the shape feature of $V_{+}$.

In the left panel of Fig.\ref{fig:4}, we plot the factorized potential $V_+(r)$ for a charged scalar perturbation in the background of a $5$-dimensional RN black hole ($\lambda = 0$). Clearly, $V_+(r)$ has one  maximum point $r=r_{\mathrm{max}}$, outside the black hole and  when $r>r_{\mathrm{max}}$, $V_+(r)$ decreases monotonically\footnote{Though  the factorized potential $V_+(r)$ is plotted here only for the case of $5$-dimensional space-time, we have actually checked that $V_+(r)$ for  massive charged scalar field in $D$-dimensional ($D\ge 4$) RN space-times share the same feature of no potential well.} as $r$ increases and its asymptotic value is just the mass of the perturbation, i.e.,  $V_+(\infty)=\mu$. Therefore, "superradiance zone" exists only when the scalar mass $\mu$ is relatively small.   

The right panel of Fig.\ref{fig:4} shows the shape of the factorized potential $V_+$ for a massless charged scalar perturbation in $5$-dimensional RN-AdS black hole space-time with different values of $\lambda$. Remarkably, there is a potential well outside the potential barrier of $V_+$ for the case $\lambda\neq0$. Nonetheless, "superradiance zone" only exists for a relatively small value of $\lambda$.\footnote{ In order to compare with the numerical results presented in previous section, we plot the factorized potential $V_+(r)$ in the left and right panel of Fig.\ref{fig:4} by taking the same values of parameters as those in Fig.\ref{fig:3} and Fig.\ref{fig:5-1}, respectively.}

Here, we would like to emphasize that  some results derived from previous numerical computation  can be also obtained at least qualitatively from analyzing the shape of factorized potential without  solving the differential equation numerically or analytically. In the following two subsections, we shall give some examples. 

\subsection{The influence of $\mu$ on the  stability of $D$-dimensional RN black hole with/without a mirror}

First,  just like those in $4$-dimensional space-time, the massive charged scalar perturbations cannot trigger superradiant instabilities in higher dimensional RN space-times without a surrounding mirror. This conclusion can be drawn directly from the shape of the factorized potential: Although the superradiant modes are allowed for massless or light massive charged scalar perturbation, there is no propagative zone sandwiched between two evanescent zones. In other words,  there is no "bound modes" if there exists no mirror.

However, if a reflecting mirror is placed outside of a RN black hole in four or higher dimensional space-time, the bound modes will appear due to the equivalence between a mirror and a infinite high potential at the same location. On the other side, as we have shown in Fig.\ref{fig:3}, the real part $\omega_R$ approaches to the mass  when the mirror moves to spatial infinity. Therefore, in the case of $r_m\rightarrow \infty$,  to make the RN black hole mirror system unstable, we should let 
\begin{equation}\label{instability condition for RN-mirror} 
	V_+(r_+)=q\Phi_H>\mu.
\end{equation} 
This is exactly the condition for the existence of a "superradiance zone".
By substituting Eqs.(\ref{func: f(r)}), (\ref{Pot:the electromagnetic potential}) and (\ref{V}) into Eq.(\ref{instability condition for RN-mirror})  and set $\lambda=0$, condition (\ref{instability condition for RN-mirror}) can be reduced to
\begin{equation}\label{eqn:q-mass-condition}
	\frac{q}{\mu}>\sqrt{\frac{2(D-3)}{D-2}}Q^{-1}=\sqrt{\frac{2(D-3)}{D-2}}r_-^{-\frac{D-3}{2}}. 
\end{equation}

When the mirror radius takes a finite value, it is difficult to derive analytically the bound of the superradiant instability regime of the composed black-hole-mirror system. To obtain the exact bound, we have to solve the two-point boundary value problem with the condition $\omega_I(r_m, \mu, q, \ell, D, r_-, r_+=1, \lambda=0)=0$. In fact, according to the shape of the factorized potential, it is pretty convenient to obtain a necessary condition for the superradiant instability. Clearly,  if the mirror is placed at the location where $V_+(r)=\omega_c$,\footnote{Actually, $r=r_+$ is always obey this constraint equation, but of course the mirror can not be placed at the event horizon, so the location which we refer to is the solution next to  $r=r_+$.} there must be no bound mode located in the "superradiance zone".

\begin{figure}
	\includegraphics[width=0.45\textwidth]{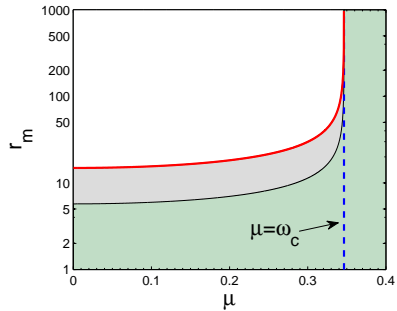}
	\caption{The $r_m\,$-$\,\mu$ plane of parameter space  for a massive charged scalar perturbation in $5$-dimensional RN black hole ($\lambda=0$) space-time, where other parameters are fixed to be $q=4, \ell=1$ and $r_-=0.1$. The red thick solid curve denotes the exact bound of the superradiant instability regime (white region), which is obtained from  our numerical result with condition $\omega_I=0$, while the thin black solid curve is estimated from the factorized potential analysis. The  vertical blue dashed line denotes the mass bound $\mu=\omega_c$.}
	\label{fig:8}     
\end{figure}

In Fig.{\ref{fig:8}}, we plot the exact bound of the superradiant instability regime as well as the one estimated according to the condition $V_+(r_{m}^{\mathrm{bound}})=\omega_c$ in the $r_m\,$-$\,\mu$ plane of parameter space where other parameters are fixed.  
In the green region, the black-hole-mirror system is stable due to no superradiance zone. In the region between the two bound curves, the system is also stable because, although there exists a small superrdiance zone in factorized potential $V_+(r)$,  the frequency of fundamental bound mode of scalar perturbation is slightly greater than critical frequency for superradiance. Obviously, the estimated bound from factorized potential analysis can serve as a necessary condition for the superradiant instability. Note that the  vertical blue dashed line in Fig.{\ref{fig:8}}  denotes the mass bound $\mu=\omega_c$, which is equivalent to Eq.(\ref{eqn:q-mass-condition}).  

\subsection{The influence of $\lambda$ on the stability of $D$-dimensional RN-AdS-black-hole-mirror system under a massless charged scalar perturbation}

\begin{figure*}
	\subfigure { \label{fig:5d}     
		\includegraphics[width=0.3\textwidth]{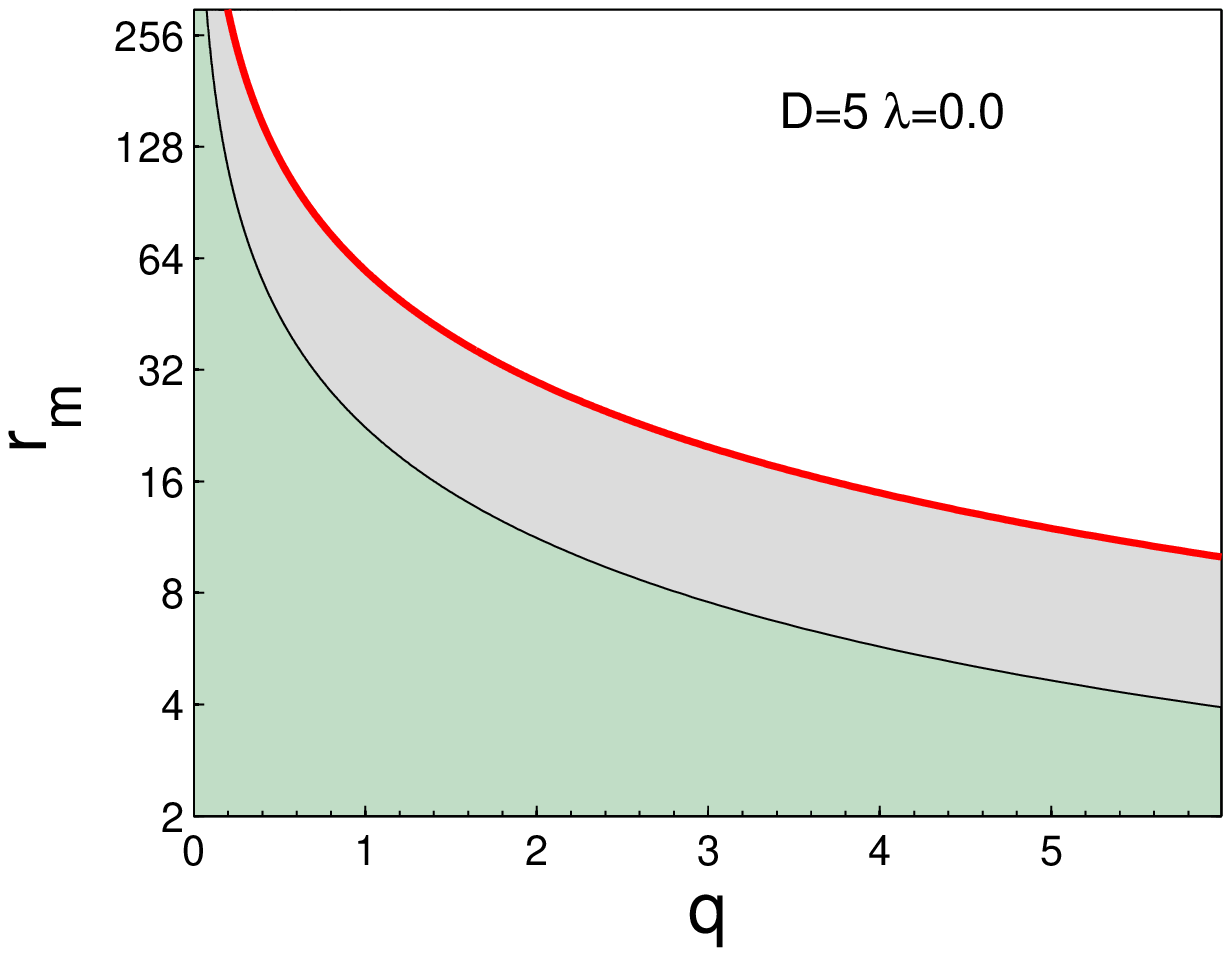} 
	}
	\subfigure { \label{fig:5e}     
		\includegraphics[width=0.3\textwidth]{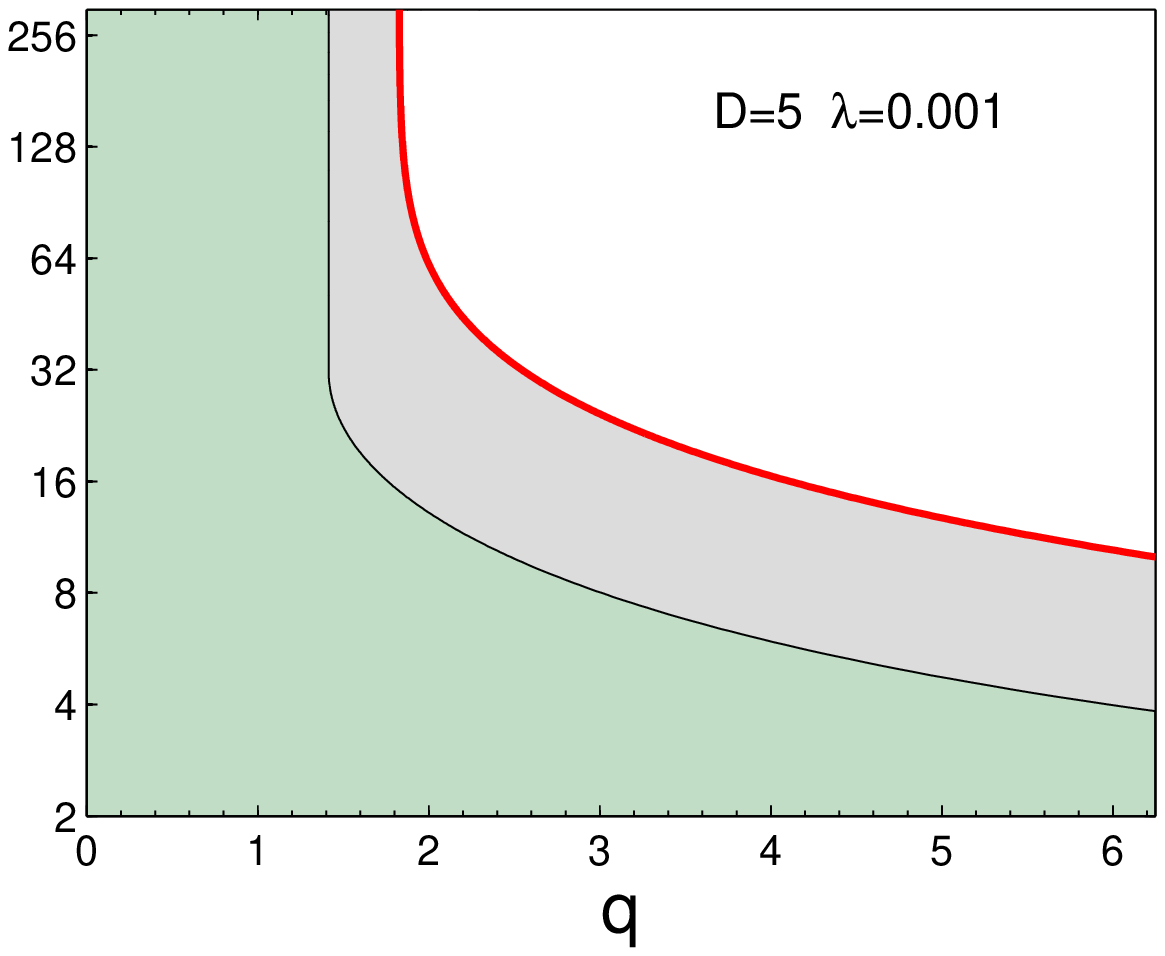} 
	}
	\subfigure { \label{fig:5f}     
		\includegraphics[width=0.3\textwidth]{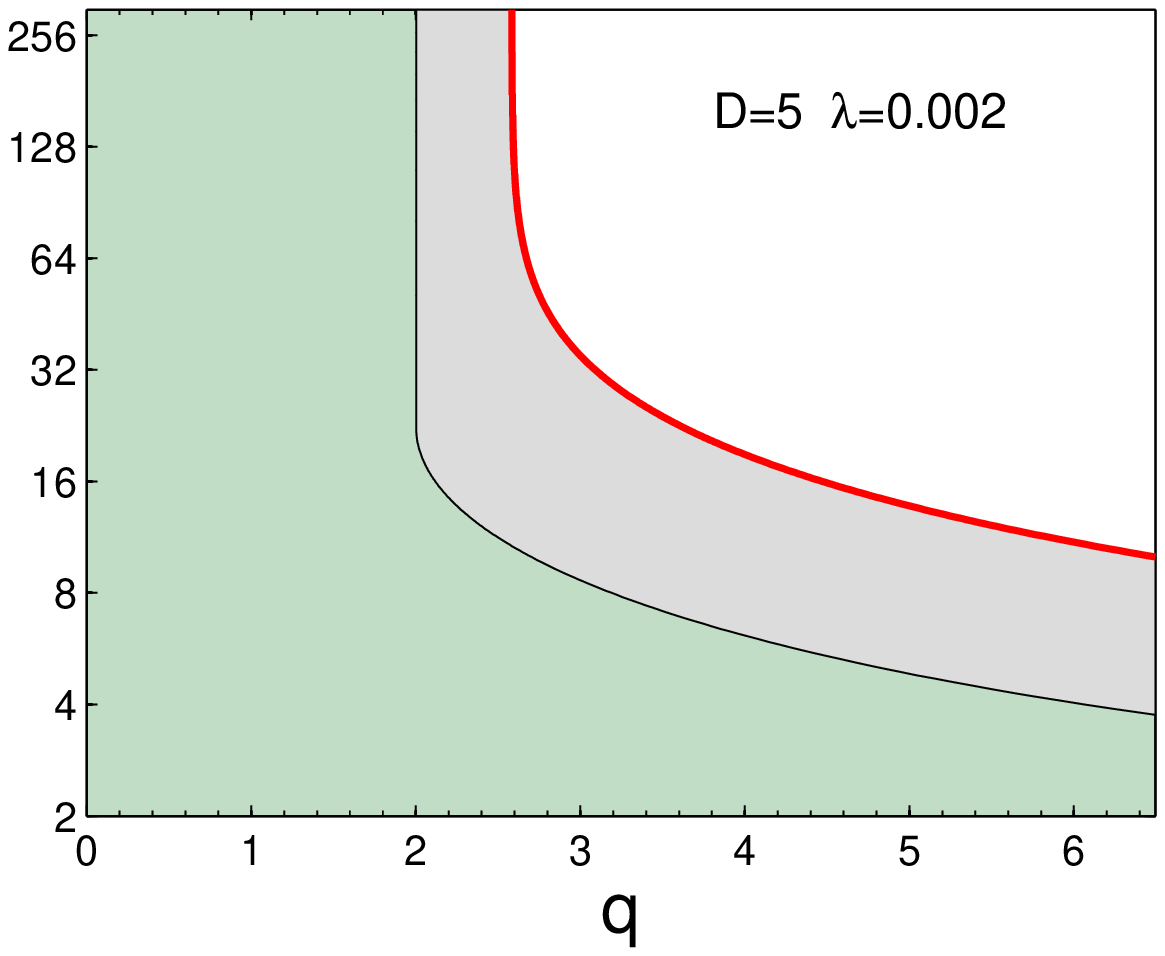} 
	}
	\caption{The $r_m\,$-$\,q$ plane of parameter space  for a massless charged scalar perturbation in $5$-dimensional RN/RN-AdS black hole space-time with different value of $\lambda$, where other parameters are fixed to be $\ell=1$ and $r_-=0.1$. The red curve denotes the exact bound of the superradiant instability regime (white region), which is obtained from  our numerical result with condition $\omega_I=0$, while the black curve is estimated from the factorized potential analysis.}
	\label{fig:5}
\end{figure*}

Directly from the shape of the factorized potential shown in the right panel of Fig.(\ref{fig:4}), we can draw the conclusion that even without a reflecting mirror, a small $D$-dimensional RN-AdS black hole may be superradiantly unstable, because when the value of $\lambda$ is small enough, the potential well can be located in the superradiance zone.\footnote{It is worth noting that the existence of potential well in superradiance zone is not a sufficient condition for superradiant instability. For example, although  when $\lambda=0.004$ there exists a "superradiance zone" in the right panel of Fig.\ref{fig:4}, the imaginary part $\omega_I$ is always negative (see the right panel of Fig.\ref{fig:5-1}).} This argument has been confirmed by our previous numerical results.

As has been shown in Fig.\ref{fig:5-1}, the real part $\omega_R$ decreases monotonically as $r_m$ increases for a massless charged scalar perturbation and its asymptotic value is given by the AdS spectrum $\omega_{\mathrm{AdS}}$. Therefore, a condition for superradiant instability in RN-AdS background is 
\begin{equation}\label{AdS condition}
	q\Phi_H>\omega_{\mathrm{AdS}}.
\end{equation}
In fact, as long as there exists one superradiantly unstable mode, the system can be said unstable. Therefore,  substituting Eq.(\ref{extrem charge}) and Eq.(\ref{AdS spectrum}) with $N=\ell=0$  into condition (\ref{AdS condition}) and using $Q<Q_c$ (equivalently, $r_-<r_+$), we obtain 
a necessary condition for superradiant instability in the limit of $r_m\rightarrow \infty$,\footnote{For RN-AdS black-hole-mirror system,  $r_m\rightarrow \infty$ is equivalent to no mirror.} 
\begin{equation}\label{AdS condition-2}
	q^2 > \frac{2(D-3)(D-1)^2}{(D-2)(\lambda^{-1}+\frac{D-1}{D-3})}.
\end{equation}
Clearly, for a given value of charge $q$, $\lambda$ should be small enough to satisfy the inequality (\ref{AdS condition-2}), that is, a small RN-AdS black hole is prone to showing superradiant instability. However, once the value of $\lambda$ holds fixed,   the black hole mirror system will be unstable only under perturbations with large value of charge $q$.   As is mentioned above,  since there is no bound on charge $q$ at least at the classical level, a large RN-AdS black hole can still be unstable provided that the value of charge $q$ is large enough.

When the mirror is not far away from the black hole,  we can  use the condition $V_+(r_{m}^{\mathrm{bound}})=\omega_c$ again to estimate the bound of superradiant instability. Of course, the exact bound still has to be obtained by integrating the differential equation (\ref{eq:wave13}) numerically.  
In Fig.{\ref{fig:5}}, we plot the exact bound of the superradiant instability regime as well as the one estimated according to the condition $V_+(r_{m}^{\mathrm{bound}})=\omega_c$ in the $r_m\,$-$\,q$ plane of parameter space where other parameters are fixed.  
Although the estimated bound is not exact, it determines the condition of the superradiant instability approximately. Furthermore, this bound can be considered as a necessary condition for the superradiant instability since the frequency of fundamental mode is greater than the critical  frequency for superradiance. 

\section{Conclusions}
In the present paper, we have performed a detailed analysis for superradiant (in)stability of a $D$-dimensional  RN-AdS-black-hole-mirror system under charged scalar perturbations. The  frequencies of the perturbation modes are obtained by numerical computation and they are in a good agreement with previous analytical and numerical results in some degenerate case. It is found that the mirror radius as well as the mass of the scalar perturbation, the cosmological constant (or equivalently AdS radius) and the dimension of space-time has important influence on the stability of the system. In order for the system to become unstable the mirror radius have to be greater than  a threshold value. By choosing a suitable value of the mirror radius, the degree of instability can be magnified by several orders of magnitude. In higher dimensional space-time, the degree of instability of the superradiant modes are severely weakened. For the higher-dimensional RN asymptotically flat space-time, the mass of the scalar perturbations alone can not give rise to superradiant instability. For a perturbation with given value of charge, a small RN-AdS black hole can be superradiantly unstable while a large one may be always stable, regardless of the existence of reflecting mirror. We find that all of the results derived from numerical computation  can be easily understood with the help of factorized potential analysis. Especially,  necessary conditions for superradiant instability can be estimated directly from the shape of the factorized potential.


It should not be denied that what we have studied in this work is of less relevance for astrophysics, but understanding the behavior of a perturbed BH surrounded by a reflecting mirror in asymptotically AdS space-time, as a first step to handle the problem of nonlinear development of the superradiance of black hole \cite{Sanchis-Gual:2015lje,Sanchis-Gual:2016tcm}, is of  interest for the researches inspired by gravity/gauge duality such as holographic superconductors \cite{Hartnoll:2009sz}, since it is easy and natural to introduce such a mirror in these dual systems.

\section*{Acknowledgments}

This work is supported in part by Science and Technology Commission of Shanghai Municipality under Grant No.~12ZR1421700 and National Natural Science Foundation of China under Grant  No.~11275128.


\end{document}